\documentstyle{article}
\begin{document}
\setlength{\baselineskip}{3.0ex}

\vspace*{2.3cm}
\setlength{\baselineskip}{3.0ex}
\begin{center}
{\large\bf Ideas on charm and beauty decays}\\
\vspace*{6.0ex}
{\large Harry J. Lipkin}\\
\vspace*{1.5ex}
{\large\it Department of Nuclear Physics, Weizmann Institute of Science}\\
{\large\it Rehovot 76100, Israel}\\
and\\
{\large\it Raymond and Beverly Sackler Faculty of Exact Sciences}\\
{\large\it School of Physics and Astronomy, Tel Aviv University, Tel Aviv,
Israel}\\

{\bf Abstract}
\end{center}

\noindent
The parton model, a zero-order approximation
in many treatments, is shown to be a ``semiclassical" model whose results
for certain
averages also hold (correspondence principle) in quantum mechanics. Algebraic
techniques developed for the M\"ossbauer effect exploit simple features of
commutators to obtain sum rules showing the validity of the
parton model for $b \rightarrow c$ semileptonic decays
in the classical limit, $\hbar \rightarrow 0$, where all
commutators vanish, and in general, even when binding effects are included,
for the lowest moments of the lepton energy spectrum at fixed
3-momentum transfer.
Interference between the $u \bar u$ and $d \bar d$
components of the $\rho^o$ and $\omega$ wave functions can be used as clues
to contributions from small weak amplitudes and CP violation in decays to
final states including these vector mesons.
\def\bra#1{\left\langle #1\right\vert}
\def\ket#1{\left\vert #1\right\rangle}
\def\VEV#1{\left\langle #1\right\rangle}
\newpage

\section{Parton model (M\"ossbauer) sum rules  for $b \rightarrow c$ decays}

We assume that bound states of one heavy quark
and other degrees of freedom are described by a Hamiltonian depending upon
the heavy quark flavour only via its mass. The dynamics of the other
degrees of freedom, including their interactions with the heavy quark,
are described by a flavour-independent operator $\Delta H$, which
depends on the heavy quark co-ordinate $\vec X$ and on the other
degrees of freedom, denoted by $\xi_\nu$, but not on the
heavy quark momentum $\vec P$. Thus $[\Delta H, \vec X] = 0$, but
$[\Delta H, \vec P] \not= 0$ and we can write the Hamiltonian $H_Q$
for systems containing a single heavy quark of flavour $Q$ = $b$ or $c$.
$$ H_Q = H(\vec P, m_Q, \vec X, \xi_\nu) = \sqrt {m_Q^2 + \vec P^2 } +
\Delta H; ~ ~ ~ ~ ~ ~ (Q=b,c)     \eqno(1.1)  $$

The hadronic transition in
semileptonic $b \rightarrow c$ decays is described by the matrix element
$ \bra{f_c}J(\vec q)\ket{i_b}$ of the
fourier component carrying three-momentum $(\vec q)$ of the flavour-changing
weak current between an initial state $\ket{i_b}$ containing
one and only one valence $b$ quark and a final state $ \ket{f_c}$ containing
one and only one valence $c$ quark. We assume that
$ J(\vec q) $ depends only on $\vec X$, normalize the current
and define moments of the final state energy distribution to obtain
$$ [J(\vec q), \Delta H]  =  0; ~ ~ ~ ~ ~ ~ ~ ~ ~
\sum_{\ket {f_c}} |\bra{f_c}J(\vec q)\ket{i_b} |^2 =
\bra{i_b} J^{\dag}(\vec q) J(\vec q)\ket{i_b} = 1
 \eqno (1.2) $$
$$ \langle {[E_c(\vec q)]^n}\rangle
\equiv \sum_{\ket {f_c}} (E_c)^n |\bra{f_c}J(\vec q)\ket{i_b}|^2
= \bra{i_b} J^{\dag}(\vec q) (H_c)^n J(\vec q)\ket{i_b} =
$$ $$ = \bra{i_b} J^{\dag}(\vec q)
\left\{ \sqrt {m_c^2 + \vec P^2 } + \Delta H \right\}^n
J(\vec q)\ket{i_b}
\eqno(1.3)  $$

These assumptions hold in
a number of conventionally used models, and in particular in the
nonrelativistic constituent quark potential models with various potentials.
Spin effects are neglected; they are taken
into account in a more detailed treatment\cite{HJLPL} .

The information about the other degrees of
freedom $\xi_\nu$ in the moments (1.3) appears
in the operator $\Delta H$ and disappears when
$ \Delta H$ acts directly either to the left or to the right on the initial
state $\ket{i_b}$.
$$  \Delta H \ket{i_b} =
(H_b -  \sqrt {m_b^2 + \vec P^2 }) \ket{i_b} =
\left(M_i -  \sqrt {m_b^2 + \vec P^2 }\right) \ket{i_b}
   \eqno (1.4) $$
where $M_i$ is the the eigenvalue of $H_b$ in the initial state $\ket{i_b}$
and is just the mass of this state.

Eqs. (1.3-1.4) express the basic physics of this approach.
In any model satisfying the assumptions (1.1-1.2)
the moments (1.3) with $n \leq 2$,
where all the $ \Delta H$ factors can be moved either
to the left or to the right so that they act on $\ket{i_b}$, are expressed as
expectation values in the initial state of single-particle operators which
depend only upon the dynamical variables of the heavy quark and are determined
completely by the one-particle density matrix for the heavy quark in the
initial state. They are the same as the
results for a naive parton model whose parton distribution
is given by this one-particle density matrix. The case $n=0$ is just
the Bjorken sum rule which effectively states that the heavy quark lifetime
is independent of binding except for phase space factors.

Only in the moments for $n \geq 3$
where commutators of the form $[\Delta H, \vec P]$ appear do
deviations from parton results occur. These are proportional to commutators
which vanish in the classical limit where $\hbar \rightarrow 0$.

Explicitly, for $n=1$ and $n\geq 2$
 $$ \langle {[E_c(\vec q)]}\rangle
= \bra{i_b} J^{\dag}(\vec q)
 \sqrt {m_c^2 + \vec P^2 }
J(\vec q) + \{M_i - \sqrt {m_b^2 + \vec P^2 } \}
J^{\dag}(\vec q) J(\vec q)\ket{i_b}
   \eqno (1.5a) $$
$$ \langle {[E_c(\vec q)]^n}\rangle
= \bra{i_b} \left(\{M_i -  \sqrt {m_b^2 + \vec P^2 }\} J^{\dag}(\vec q) +
J^{\dag}(\vec q)  \sqrt {m_c^2 + \vec P^2 }\right)
(H_c)^{(n-2)}\cdot $$
$$ \cdot\left( J(\vec q)\{M_i -  \sqrt {m_b^2 + \vec P^2 }\} +
 \sqrt {m_c^2 + \vec P^2 }\cdot J(\vec q)\right) \ket{i_b}
   \eqno (1.5b) $$
The sum rules for for  n=1 and n=2 are simplified\cite{HJLPL}
by expressing $J(\vec q)$ explicitly in terms of $\vec X$,
the heavy quark mass difference $\delta m \equiv m_b - m_c$, and the free
recoil energy $R(\vec q)$ and the ``isomer" or ``isotope" shift $I_{bc}$,
defined respectively as
$$
R(\vec q) \equiv
H [(\vec P + \vec q), m_c, \vec X, \xi_\nu]  -
H [(\vec P ), m_c, \vec X, \xi_\nu] =
\sqrt {(\vec P + \vec q)^2 + m_c^2} - \sqrt {\vec P^2 + m_c^2}
\approx {{q^2} \over {2 m_c}}
   \eqno (1.6a) $$
$$
I_{bc} \equiv  \delta m +
H [\vec P , m_c, \vec X, \xi_\nu]  -
H [\vec P , m_b, \vec X, \xi_\nu] \approx \vec P^2 \cdot
{{\delta m} \over {2 m_c m_b}}
   \eqno (1.6b) $$
$$ \langle {[E_c(\vec q)]^n}\rangle=
\bra{i_b}\{M_i + R(\vec q) + I_{bc} - \delta m
\}^n  \ket{i_b}     \eqno (1.6c) $$
where $\approx$ denotes the nonrelativistic approximation.
These sum rules can also be written for the energy $E_W$ carried by
the W; i.e. by the leptons,
$$ \langle {E_W(\vec q)}\rangle \equiv
\sum_{\ket {f_c}} E_W
|\bra{f_c}J(\vec q)\ket{i_b}|^2
= M_i - \langle {[E_c(\vec q)]}\rangle
= \delta m - \bra{i_b}R(\vec q) + I_{bc}\ket{i_b}
$$ $$ \approx
\delta m - {{q^2} \over {2 m_c}} -
\bra{i_b}\vec P^2\ket{i_b}\cdot {{\delta m} \over {2 m_c m_b}}
   \eqno (1.7a) $$
$$ \langle {[E_W(\vec q)]^2}\rangle - \langle {[E_W(\vec q)]}\rangle^2 =
\bra{i_b}\{R(\vec q) + I_{bc}\}^2\ket{i_b} -
\bra{i_b}\{R(\vec q) + I_{bc}\}\ket{i_b}^2 \approx
$$ $$ \approx
{{\bra{i_b} \vec P^2 \ket{i_b} \cdot q^2 }\over{3m_c^2}} +
{{(\delta m)^2} \over {4 m_c^2 m_b^2}} \cdot
(\bra{i_b} P^4\ket{i_b} - \bra{i_b} P^2\ket{i_b}^2)
    \eqno (1.7b) $$
An upper bound for the transition to a given final state $\ket{f_m}$ with
energy $ E_m$ is obtained by replacing all
energies except $E_m$ in the sum rule with the lowest possible energy
$E_g = M_D + {{q^2} \over {2 M_D }}$; the energy of the
lowest available state of the charmed system,
$$ E_m |\bra{f_m}J(\vec q)\ket{i_b}|^2 + E_g(1-|\bra{f_m}J(\vec q)\ket{i_b}|^2)
\leq \langle {[E_c(\vec q)]}\rangle
= M_i + R(\vec q) + I_{bc} - \delta m $$ $$ \approx
M_i  + {{q^2} \over {2 m_c}} +
\bra{i_b}\vec P^2\ket{i_b}\cdot {{\delta m} \over {2 m_c m_b}}- \delta m
   \eqno (1.8a) $$
$$
|\bra{f_m}J(\vec q)\ket{i_b}|^2) \leq
 {{\langle {[E_c(\vec q)]}\rangle - E_g }\over{E_m - E_g}} \approx
{{1}\over {E_m - E_g}} \cdot\left(
 {{q^2} \over {2 M_D m_c}}\cdot [ M_D - m_c] +  \epsilon
\right)
   \eqno (1.8b) $$
where
$$   \epsilon \equiv [M_i - m_b] - [ M_D - m_c] +
\bra{i_b}\vec P^2\ket{i_b}\cdot {{\delta m} \over {2 m_c m_b}}
=  \bra{i_b} H_c \ket{i_b} - M_D  \eqno (1.8c) $$
The matrix element $ \bra{i_b} H_c \ket{i_b}$ gives
a value for $M_D$ exact to first
order in the perturbation $H_c - H_b$ and in the reciprocal
mass difference ${{m_b - m_c}\over{m_b m_c}}$. Thus $\epsilon$ is second order
in $1/m_c$.

Thus the probability of excitation by an energy $E_m -E_g$ is bounded by the
ratio to this energy of the small energy $ {{q^2} \over {2 M_D m_c}}\cdot
[ M_D - m_c]$ which goes to zero as $ q^2 \rightarrow 0$ with a small
correction
$\epsilon$ which vanishes in the heavy quark symmetry limit.
This treatment can be extended to include spin and relativistic effects.
However it can be expected to be already particularly good in the low-recoil
domain of small $q^2$ where the bound (1.8) places serious limits on the
probability of high excitations; i.e. on low lepton energies.

\section{Use of $\rho - \omega$ Interference Effects
in B and D Decays}

The $\rho^o$ and $\omega$ are equal mixtures with opposite relative phase of
the $u \bar u$ and $d \bar d$ vector quarkonium states, which we denote
respectively by $V_u$ and $V_d$. Thus if $\rho^o$ and
$\omega$ are produced via a quark diagram leading to a single flavour
state, either $V_u$ or $V_d$, they should both be produced
equally\cite{ALS}  with a definite relative phase and show the
interference effect originally suggested by Glashow
\cite{GLASHOW} and subsequently extensively observed experimentally
\cite{GOLDHABER}. These predictions are
particularly sensitive via interference to small contributions from
other diagrams producing the state forbidden in the dominant
diagram\cite{PKEKETA}. Similar effects have been considered for decays into
$\eta - \eta'$ modes \cite{PKETA}.

The relation between production and decay processes of the $\rho^o$ and
$\omega$
can be understood by a comparison with the $K - \bar K$ system. The four
``quark-flavour" vector meson eigenstates $\rho^+ (u \bar d)$, $V_d (d \bar
d)$,
$V_u (u \bar u)$ and $\rho^-(d \bar u)$ are directly analogous to the four kaon
quark-flavour eigenstates:  $K^+ (u \bar s)$,  $K^o (d \bar s)$,
$\bar K^o (s \bar d)$ and  $K^- (s \bar u)$. In both cases the two neutral
states are nearly degenerate and both quark-flavour eigenstates can decay into
two pions or into three pions.

The decay interaction mixes the quark flavour
eigenstates into short-lived mesons $K_S$ and $V_S$ (or
$\rho^o$), which decay dominantly into two pions, and long-lived mesons
$K_L$ and $V_L$ (or $\omega$), which decay dominantly into three pions.
The decay eigenstates are both
eigenstates to a very good approximation of a symmetry, CP for the kaons and
G-parity for the vectors, which forbids the $2\pi$ decay for $K_L$ and
$\omega$. However because both $CP$ and $G$ are
broken by relatively small effects both the $K_L$ and $\omega$ have
a small $2\pi$ branching ratio and interesting interference effects are
observed.

Neutral kaons are produced and leave the production vertex as flavour
eigenstates $K^o$ and $\bar K^o$. They decay
after leaving the range of all final state interactions
as equal mixtures of $K_L$ and $K_S$ with opposite relative phases.
If neutral vector mesons are similarly produced as flavour eigenstates
and decay only after leaving the range of all final state interactions, they
decay as $V_u$ and $V_d$; i.e as equal mixtures of $\rho^o$ and $\omega$ with
opposite relative phases. This leads to interesting experimental consequences
which can be useful for investigations of weak interactions and $CP$ violation.
However the lifetimes here are much shorter and the escape from
the range of final state interactions before decay is open to question.

Good experimental evidence that the vector mesons do decay
outside the range of final state interactions was first noted in strong
interaction reactions described by diagrams where a final state with one
vector flavour eigenstate is forbidden by the Alexander-Zweig
\cite{ALS} or OZI rule; giving a selection rule and
predicting the equality of the two observable cross sections,
$$ \sigma (K^-p  \rightarrow \Lambda V_{d})=0; ~ ~ ~ ~ ~ ~ ~ ~ ~ ~ ~ ~
\sigma (K^-p  \rightarrow \Lambda \omega) =
\sigma (K^-p  \rightarrow \Lambda \rho^o) \eqno(2.1) $$
This prediction from the implied $\rho^o$ and $\omega$ production
via $V_u$ was confirmed by experiment and the $\rho-\omega$
interference subsequently observed
\cite{CERN}. Final state interactions are expected to be very
different for the $\Lambda \omega$ and $\Lambda \rho^o$ states since they
have different isospins and are coupled to completely different hadronic
channels. Thus the experimentally observed equality (2.1) is
evidence that the decay occurs outside the range of final state interactions.

In weak interactions there are a number of ways to test whether the decay
occurs outside the range of final state interactions. Final state
interactions should be directly observed in final states involving $\rho$
mesons as a perturbation of the Breit-Wigner
shape of the decay pion spectrum. In semileptonic decays to $\rho^o$ and
$\omega$ where there are no final state interactions, the $\rho^o$ and $\omega$
should be produced equally from the produced flavour eigenstate, with
a relative phase measurable by interference.

Weak interaction diagrams tend to producd the $\rho$ and $\omega$ via
only their $V_{d} $ or $V_{u}$ components since the quark lines in these
diagrams have definite flavour labels. In the $B^+ \rightarrow K^+
\rho^o$ and $B^+ \rightarrow K^+ \omega$ decays $K^+ V_u$ is produced by both
the Cabibbo-suppressed color-favored
and Cabibbo-suppressed color-suppressed spectator tree diagrams
and also by other diagrams like the penguin which first produce a $\bar s u$
intermediate state and then produce the additional $q \bar q$ via gluons
  $$ B^+(\bar b u) \rightarrow _{(cstree)}  \rightarrow (\bar u u \bar s) u
  \rightarrow K^+ V_u; ~ ~ ~ ~ ~ ~ ~   B^+(\bar b u)
\rightarrow _{(penguin)} \rightarrow \bar s u  \rightarrow
  K^+ V_u \eqno(2.2)$$
Production of $K^+ V_d$ is OZI forbidden both for the penguin diagram (2.2) and
for diagrams producing a $\bar d d$ pair by final state interactions following
the tree diagram. The OZI rule forbids all processes where both members of a
$\bar q q$ pair produced by gluons end up in the same final state hadron. In
this particular case the production of $K^+ V_d$ is also forbidden by flavour
SU(3), even without assuming OZI, for all transitions via an intermediate
$\bar s u$ state. A spin-zero $K^+ V_d$ state has exotic flavour quantum
numbers
and cannot couple to a single quark-antiquark pair. This most easily seen by
noting the exotic flavour quantum numbers $J^{PG} = 0^{++}$ of the
$\pi^+ \phi$ state related to $K^+ V_d$ by the SU(3) transformation which
exchanges $d$ and $s$ flavours.
By analogy with (2.1) we obtain
$$ BR(B^+  \rightarrow K^+ V_{d})=0 ~ ~ ~ ~ ~ ~ ~ ~ ~ ~ ~ ~
BR(B^+  \rightarrow K^+ \omega) =
BR(B^+  \rightarrow K^+ \rho) \eqno(2.3) $$
This prediction can be checked directly by experiment and the same
$\rho-\omega$
interference observed in the strong reaction (2.1)\cite{CERN} should also be
observed here.

The interference is observable in detailed analysis of the
$\pi^+ \pi^-$ spectrum over the mass range of the $\rho$ resonance.
The isospin violating $\omega \rightarrow \pi^+ \pi^-$ has a branching
ratio of 2.2\%. The width of the $\omega$ is 8.4 MeV while that of the
$\rho$ is 149 MeV. Thus if the $\rho$ and $\omega$ are produced equally
in any reaction or decay, the $\pi^+ \pi^-$ decay mode seen at the omega
peak will come from both the $\rho$ and the $\omega$ and the relative
intensities of the two contributions is given by:
$$ {{I_\omega(B\rightarrow \pi^+ \pi^- X)}\over
{I_\rho(B\rightarrow \pi^+ \pi^- X)}}
\approx 0.022 \cdot {149 \over 8.4} \approx 0.39
\eqno(2.4) $$
If the two contributions are coherent, the total contribution is given by
$$ {{I_{total}(B\rightarrow \pi^+ \pi^- X)}\over{
I_\rho(B\rightarrow \pi^+ \pi^- X)}} \approx
(1 + \sqrt {0.39} \cos \alpha)^2=
1 + 1.25 \cos \alpha + 0.39 \cos^2 \alpha \eqno(2.5) $$
where $\alpha$ is the relative phase of the $\rho$ and $\omega$
contributions.

If these predictions are confirmed experimentally, the same
approach can be used for the more interesting case of
$B^o \rightarrow K^o \rho^o$ and $B^o \rightarrow K^o \omega$ decays, where
the Cabibbo-suppressed color-suppressed spectator tree diagram again
produces $V_u$
but the penguin diagram and all other diagrams which go via an intermediate
$\bar q q$ pair produce $V_d$.
Tree production of $K^o V_d$ and penguin production of $K^o V_u$ are both
OZI and SU(3) forbidden. Thus
  $$ B^o(\bar b d) \rightarrow _{(cstree)}  \rightarrow (\bar u u \bar s) d
  \rightarrow K^o V_u; ~ ~ ~ ~ ~ ~ ~ ~ B^o(\bar b d) \rightarrow _{(penguin)}
\rightarrow \bar s d  \rightarrow
  K^o V_d \eqno(2.6)$$
   $$ {{BR (B^o \rightarrow K^o \rho^o)}\over{
BR (B^o \rightarrow  K^o \omega )}}=
\left| {{T + P}\over { T-P}}\right| ^2 =
\left| 1 + {{2P}\over { T-P}}\right| ^2
\approx 1 + 4 Re(P/T)
 \eqno(2.7a) $$
   $$ {{BR (\bar B^o \rightarrow \bar K^o \rho^o)}\over{
BR (\bar B^o \rightarrow \bar K^o \omega )}}=
\left| {{\bar T + \bar P}\over { \bar T-\bar P}}\right| ^2 =
\left| 1 + {{2\bar P}\over { \bar T-\bar P}}\right| ^2
\approx 1 + 4 Re(\bar P/\bar T)
 \eqno(2.7b) $$
where $T$, $P$, $\bar T$ and $\bar P$  denote respectively the contributions
to the
decay amplitudes (2.7a) and to the charge conjugate decay amplitudes (2.7b)
from tree and penguin diagrams.

This offers the possibility of detecting the penguin contribution and
also measuring the relative phase of penguin and tree contributions, as
well as detecting $CP$ violation in a difference between the
charge-conjugate $\rho/\omega$ ratios (2.7a) and (2.7b).
The relations (2.7) provide additional
input from $B \rightarrow K \omega$ decays that can be combined with
isospin analyses of $B \rightarrow K \rho$ decays to
separate penguin and tree contributions \cite{PBPENG}.
A similar additional input is obtainable from combining $\omega$
decay modes with isospin analyses of other $\rho$ decay
modes\cite{GRONAU}

In the
Cabibbo-favored $B^o \rightarrow \bar D^o \rho^o$ and $B^o \rightarrow \bar
D^o \omega$ decays, the color-suppressed spectator tree diagram produces $V_d$
but the W-exchange diagram and all other diagrams which go via an intermediate
$\bar c u$ pair now produce $V_u$ in the transitions allowed by OZI,
  $$ B^o(\bar b d) \rightarrow _{(cstree)}  \rightarrow (\bar c u \bar d) d
\rightarrow \bar D^o V_d; ~ ~ ~ ~ ~ ~ ~ ~ ~
B^o(\bar b d) \rightarrow _{(Wexc)}
\rightarrow \bar c u  \rightarrow \bar D^o V_d \eqno(2.8)$$
   $$ {{BR (B^o \rightarrow \bar D^o \rho^o)}\over{
BR (B^o \rightarrow  \bar D^o \omega )}}=
\left| {{T + W}\over { T-W}}\right| ^2 =
\left| 1 + {{2W}\over { T-W}}\right| ^2
\approx 1 + 4 Re(W/T)
 \eqno(2.9) $$
where T, and W denote contributions to decay amplitudes from
tree and W-exchange diagrams respectively.
Here both tree and W-exchange involve the same combination of
CKM matrix elements. Thus no CP-violating relative phase is expected.

In Cabibbo suppressed decays into charmonium and $\rho$ or
$\omega$ e.g.
$B^o \rightarrow \psi \rho^o$ and $B^o \rightarrow \psi \omega$,
the color-suppressed spectator tree diagram produces $V_d$,
but the W-exchange diagram and all other diagrams which go via an intermediate
$\bar c c$ pair cannot produce a single charmonium state in transitions
allowed by OZI. Thus only the tree can contribute and
  $$ B^o(\bar b d) \rightarrow (\bar c c \bar d) d
  \rightarrow \psi V_d; ~ ~ ~ ~ ~ ~ ~ ~ ~
BR (B^o \rightarrow \psi \rho^o) =
BR (B^o \rightarrow  \psi \omega )
 \eqno(2.10) $$
and the definite relative phase for production via $V_d$ is predicted for
$\rho-\omega$ interference.

One interesting case where data are already available
\cite{PDG} is in the charm decays
$$ BR[D^o(c \bar u) \rightarrow \bar K^o \rho^o]=(6.1 \pm 3.0 ) \times 10^{-3}
; ~ ~ ~
BR[D^o(c \bar u) \rightarrow \bar K^o \omega = (2.5  \pm 0.5)\%
\eqno(2.11a)$$
$$ BR[D^o(c \bar u) \rightarrow \bar K^o \phi] = (8.8 \pm 1.2) \times 10^{-3}
\eqno(2.11b)$$
The dominant diagram for the decays (2.11) is expected to be
the color-suppressed spectator tree diagram
which gives the $V_u$ component of the $\rho$ and $\omega$,
and predicts equal branching ratios for the two decays (2.11a) and zero
for the $\phi$.
  $$ D^o(c \bar u) \rightarrow _{(cstree)}
\rightarrow (s u \bar d) \bar u \rightarrow \bar K^o
V_u                                          \eqno(2.12a)$$
The data therefore indicate the presence of another contribution.

The decay can also proceed via a color-favored spectator tree diagram
followed by a final state charge exchange rescattering
via the intermediate state $K^- \rho^+$.
The final state interaction can be expected to be enhanced if there is a
$K^*$ resonance in this mass region. The $\rho^o$
and $\omega$ are produced via the $V_d$ component and the $\phi$ can also
be produced.
  $$ D^o(c \bar u) \rightarrow _{(cftree)}
\rightarrow (s u \bar d) \bar u \rightarrow \bar K^-
\rho^+ \rightarrow \bar K^{*o} (s \bar d) \rightarrow \bar K^o
V_d                                          \eqno(2.12b)$$
  $$ D^o(c \bar u) \rightarrow _{(cftree)}
\rightarrow (s u \bar d) \bar u \rightarrow \bar K^-
\rho^+ \rightarrow \bar K^{*o} (s \bar d) \rightarrow \bar K^o
\phi \eqno(2.12c)$$
The $V_u$ component is not produced by this mechanism since the decay
$ \bar K^{*o} (s \bar d) \rightarrow \bar K^o V_u$ is OZI and SU(3) forbidden.

In all cases the $V_u$ or $V_d$ can hadronize into $\rho^o$ or $\omega$
and equal magnitudes are predicted for the $\rho$ and $\omega$ final
states. However, the relative phase is opposite in the two cases.
The measurement of this phase in interference experiments
can distinguish between the two mechanisms (2.12a) and (2.12b) and
check the validity of assumptions regarding color suppression in
tree diagrams and the role of final state interactions.
There should be significant peaks and dips observed in the $\pi^+ \pi^-$
spectrum in heavy meson decay modes involving the $\rho$ and $\omega$.

The $\pi^+ \pi^-$ spectrum for the color-suppressed spectator tree
contribution (2.12a) where the $\rho$ and $\omega$ are produced
via the $u \bar u$ component should be similar to that observed
\cite{CERN}
in strangeness exchange reactions with $K^-$ beams like (2.1).
If, however, the color-favored transition via an intermediate resonance
(2.12b) is dominant the interference should have the exact opposite sign.
The same interference should be observed in the Cabibbo-suppressed decay
(2.12d) which also goes via $V_d$.

Other contributions to the decays (2.12) with very different flavour
properties are
  $$ D^o(c \bar u)  \rightarrow _{(Wexc)}
\rightarrow s \bar d \rightarrow \bar K^o
V_d; ~ ~ ~ ~ ~ ~ ~ ~ ~ D^+(c \bar d)  \rightarrow _{(Ann)}
\rightarrow u \bar d \rightarrow \pi^+ \rho
                                  \eqno(2.13)$$
The W-exchange decay to $\bar K^o V_u$ is OZI and SU(3) forbidden and the
simple annihilation diagram without gluon emission from the
initial state is forbidden for the $\pi^+ \omega$ state by G-parity.
Note that the $\pi^+ \omega$ final state has the exotic quantum numbers
$J^{PG}=0^{-+}$ and cannot be produced by strong interactions from an
intermediate state containing only a single quark-antiquark pair.

We thank the Institute for Nuclear Theory at the University of Washington
for its hospitality and the Department of Energy for partial support during
the completion of this work.

\end{document}